\documentclass[twocolumn,showpacs,preprintnumbers,amsmath,amssymb,prl]{revtex4}

\usepackage{graphicx}

\begin{document}


\title{Free-energy transition 
in a gas of non-interacting nonlinear wave-particles}

\author{A.  Fratalocchi$^{1,2}$, C. Conti$^{1,2}$, G. Ruocco$^{2,3}$, S. Trillo$^{2,4}$}
\affiliation{  $^1$ Centro Studi  e
Ricerche Enrico Fermi, Via Panisperna 89/A, I-00184, Roma, Italy\\
$^2$ Research center Soft INFM-CNR, c/o University of Rome Sapienza,
I-00185, Roma, Italy\\
$^3$ Department of Physics, University of Rome Sapienza, I-00185, Rome, Italy\\
$^4$ Department of Engineering, University of Ferrara, Via Saragat 1, 44100 Ferrara, Italy 
}

\date{\today}

\begin{abstract}
We investigate the dynamics of a gas of non-interacting particle-like soliton waves, demonstrating that phase transitions originate from their collective behavior.
This is predicted by  solving exactly the nonlinear equations and by employing methods of the statistical mechanics of chaos. 
In particular, we show that a suitable free energy undergoes a metamorphosis as the input excitation is increased, 
thereby developing a first order phase transition whose measurable manifestation is the formation of shock waves. 
This demonstrates that even the simplest phase-space dynamics, involving independent (uncoupled) degrees of freedom, 
can sustain critical phenomena. 
\end{abstract}



\maketitle

The methods of statistical mechanics have permeated physics as a whole including modern areas of 
deterministic chaos \cite{TOCS}, complexity \cite{SGTAB}, and nonlinear waves \cite{KS75,Rasmussen00,bf_solg_ela}.
Among its basic notions, it is well known that  the thermodynamics associated  to a gas of non-interacting
particles  is trivial,  being characterized  by a constant free-energy with respect to control parameters. 
Conversely, in order to have cooperative phenomena,
such as phase-transitions or more complex processes,
the intervention of some sort of interaction potential must be called for.
\newline \indent
Within the context of nonlinear wave propagation,
a challenge to the generality of this picture could come from \emph{solitons},
i.e. exact solutions of integrable nonlinear partial differential equations (PDE), 
which display well-known particle-like behavior.
Stemming from the numerical experiments of Fermi-Pasta-Ulam (FPU) on the equipartition of energy in nonlinear chains \cite{CPOEF}, solitons have found applications in area as different as Bose-Einstein 
condensation (BEC) \cite{BEC}, nonlinear  optics  \cite{Kivshar98,OSFFTPC},  
fluid-dynamics \cite{LANW},  solid state  physics  \cite{SICMP}, general relativity  \cite{GS}
and many others. While relying on large (non-perturbative) nonlinear
effects, solitons conserve their number and spectral parameters (eigenvalues)
evolving without any practical interaction, albeit some displacements in their collisions.
The isospectrality allows for reducing the infinito-dimensional phase space associated to the global wave-function
to a simple one where $N$ independent degrees of freedom
corresponding to $N$ soliton particles are effective.
On this basis one can argue whether the statistical description of an ensemble of \emph{non-interacting} solitons
is actually as trivial (i.e., with a constant free energy and no critical phenomena) as inferred from the statistical mechanics of free particles, 
or else exhibits cooperative phenomena as the number of particles (degrees of freedom) grows.
In this Letter, we address this issue by investigating the statistical mechanics of a gas of soliton particles. 
By exploiting ideas from the thermostatistics of chaos \cite{TOCS}, we demonstrate that a suitably defined free-energy undergoes a metamorphosis as the
number of solitons grows, changing from a constant to a function that supports a first-order phase transition. 
Correspondingly, the overall wave-function develops a steep front (shock), which results from the
scaling properties of the soliton velocity distribution, in turn determined by the input excitation. 
Hence the shock formation can be interpreted as a cooperative process of several solitons.  
Therefore even the simplest phase-space dynamics, where particle-like waves evolve independently from the others, 
could result into phase-transitions.\\   
\indent We make reference to a universal  integrable model, namely
the defocusing nonlinear Schr\"odinger equation:
\begin{equation}
  \label{nlsd0}
  i\frac{\partial\psi}{\partial z}-\frac{\partial^2\psi}{\partial x^2}+2(\lvert\psi\rvert^2-\rho^2)\psi=0,
\end{equation}
and by means of the inverse scattering transform (IST) \cite{HMITTOS}, we calculate the field evolution after the so-called {\em dark}
initial condition $\psi(x,0)=-\rho\tanh{x}$, with $\rho$ integer. The latter is experimentally accessible and particularly important
in BEC  and nonlinear optics \cite{BEC,Kivshar98,OSFFTPC}, where the occurrence of dispersive shock waves
has been recently observed and discussed \cite{sw_so_ghofra,fleisher07,sw_ds_hoefe,PhysRevA.69.043610,PhysRevA.69.063605,el:053813}, 
though in regimes not involving dark disturbances and the soliton-driven wave-breaking discussed here.
The constraint $\rho$=integer, corresponding to a \emph{reflectionless} potential, 
allows us to investigate a novel scenario where the dispersive shock dynamics is determined solely by solitons that are embedded in the input,
with radiation waves playing absolutely no role. Unlike previous approaches to shock waves (hydrodynamic limit, Whitham averaging) 
involving different degrees of approximation \cite{we_ub_gurev,w2_ub_gurev,mt_op_kamcha,nls_ub_kamcha,sw_ds_hoefe}, 
we derive exact solutions of Eq. (\ref{nlsd0}), which enable us to contrast the smooth dynamics determined
by a small number $N$ of solitons with the strongly nonlinear case ($\rho, N \gg 1$). In the latter regime, this approach allows us to introduce
a free energy that reflects the spectral distribution of solitons in the (invariant) eigenvalue space, using a non-canonical measure \cite{TOCS}.
In this respect, our approach shows no similarities with the Gibbsian statistical mechanics of soliton-bearing systems
where nonlinear excitations are thermally controlled \cite{KS75,Rasmussen00}, 
or with the kinetic theory of random distributions of colliding solitons \cite{bf_solg_ela}. Our formulation is general and applies to every integrable system, such as integrable versions of FPU.

\paragraph{Spectral transform. ---}
The solution of Eq. (\ref{nlsd0}) is derived by means of IST in two steps:
(i) by calculating the spectral transform of the system; (ii) by solving the inverse scattering problem. It is worth to remark that the solution of Eq. (\ref{nlsd0}) with initial value $\psi_0=\psi(x,0)$ represents a far more general problem than the calculation of the generic $N-$soliton solution of Eq. (1). The former, in fact, calls for the evaluation of the spectral transform $\mathcal{S}\{\psi_0\}$ of $\psi_0$ and the solution of the inverse problem for $\mathcal{S}\{\psi_0\}$, while the latter needs just the calculation of the inverse problem for an unknown $\psi$ whose spectrum contains only solitons.
In order to highlight the main physical results, we defer unessential mathematical details to an extended publication, while referring to Ref. \cite{HMITTOS} for the IST method. 
The spectral transform $\mathcal{S}\{\psi_0\}$ is defined as:
\begin{equation}
  \mathcal{S}\{\psi_0\}=\{-\infty <\lambda<\infty, \frac{b(\lambda)}{a(\lambda)}; \lambda_n, c_n, n=1,\dotsc,N \},
\end{equation}
with the reflection coefficient $b/a=R$, $\lambda_n$ corresponding to simple zeros of $a$ (for Im$(\lambda)\geq 0$) and $c_n=b(\lambda_n)/\dot{a}(\lambda_n)$, being $\lambda$ a spectral parameter and $a(\lambda)$, $b(\lambda)$ transition coefficients relating the Jost solutions of the direct scattering problem \cite{HMITTOS}. We calculate $a$ and $b$ corresponding to $\psi_0$, obtaining:
\begin{align}
  \label{rc0}
  &a=i\frac{k\cdot\Gamma(\frac{k}{2i})^2}{\lambda\cdot\Gamma(\frac{k}{2i}+\rho)\Gamma(\frac{k}{2i}-\rho)}, &b=\frac{k\cdot\lvert\Gamma(i\frac{k}{2})\rvert^2}{2\Gamma(1-\rho)\Gamma(\rho)}\text{.}
\end{align}

\begin{figure}
\includegraphics[width=8.5 cm]{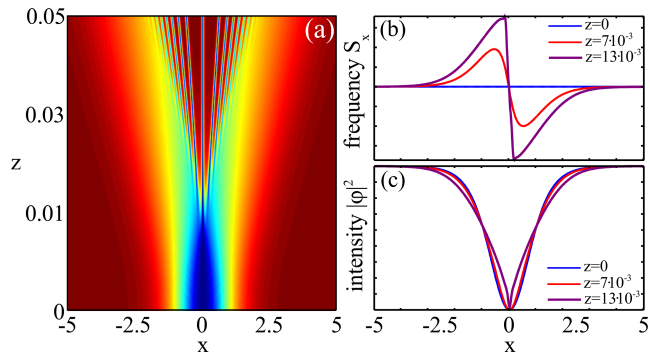}
\caption{\label{ss0}  (Color  online). 
(a) Level plot of intensity $|\psi|^2$,
and (b) snapshots of frequency $S_x$ and (c) intensity $\lvert \psi \rvert^2$, 
for input $\psi_0=30\tanh{x}$.}
\end{figure}

When $\rho$ is integer, $R=0$ and $\psi_0$ is a reflectionless potential containing $N_\rho \equiv N(\rho)=2\rho-1$ solitons. 
The spectral transform $\mathcal{S}\{\psi_0\}$ then reads:
\begin{align}
\label{stt}
& \mathcal{S}\{\psi_0\}=\{R(\lambda)=0;\lambda_{\pm n}=\pm 2\sqrt{\rho^2-(\rho-n)^2},  \nonumber\\
&c_{\pm n}=(-1)^{n+1}\frac{\prod_j(s_{\pm n}-s_j^*)}{\prod_{j\neq \pm n}(s_{\pm n}-s_j)}, n=0,\dotsc,M \},
\end{align}
with $s_n=\lambda_n+2i(\rho-n)$, $j\in[-M,M]$ and $M \equiv \rho-1$. 
\paragraph{Inverse scattering problem. ---}
The $z$-evolution of the wave function $\psi$ is found from the solution of the inverse problem \cite{HMITTOS} and, for the spectral transform represented by Eq. (\ref{stt}), reads:
\begin{align}
\label{solnd}
&\psi(x,z)=\rho\left[1+\displaystyle\frac{\lvert B\rvert}{\lvert A\rvert}\right], &B=\begin{bmatrix}
A &e\\
d &0
\end{bmatrix}, 
\end{align}
where $d_l=i\sqrt{c_l/s_l}e^{\nu_l(x-\lambda_l z)/2}$, $e_j=i2\rho d_j/s_j$, $j,l\in[-M,M]$, $\nu_j \equiv 2(\rho-j)$
and $A$ is the matrix
\begin{align}
\label{solndA}
  &A_{jl}=\delta_{jl}+i\frac{2\rho\sqrt{c_jc_l}e^{[x(\nu_l+\nu_j)-z(\nu_l\lambda_l+\nu_j\lambda_j)]/2}}{\sqrt{s_js_l}(s_j-s_l^*)}.
\end{align}
Equations (\ref{stt}) and (\ref{solnd}) describe the soliton evolution corresponding to the input $\psi_0$.
Note that $\lambda_n$ represent the invariant velocity of the $n$-th soliton. 
\begin{figure}
\includegraphics[width=8.5 cm]{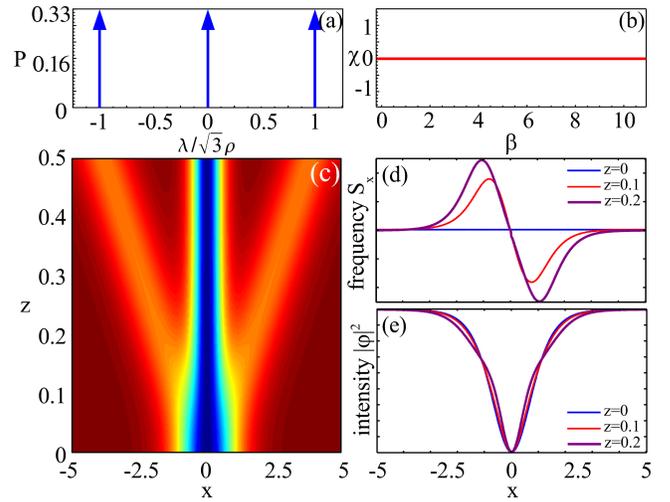}
\caption{\label{twd} (Color  online) 3-dark soliton solution corresponding to $\psi_0=2 \tanh{x}$: 
(a) Eigenvalue measure $P$ ($p_j/\varepsilon$ for $\epsilon\rightarrow 0$); 
(b) numerically calculated free energy $\mathcal{X}$ vs. $\beta$; 
(c) Pseudo-color level plot of intensity $|\psi|^2$;
(d)-(e) Profiles of frequency $S_x$ and intensity 
$|\psi|^2$ at different $z$ .}
\end{figure}
\paragraph{Soliton-gas thermodynamics. ---}
To derive a soliton-gas thermodynamics, we begin by decomposing the spectral transform $\mathcal{S}\{\psi_0\}$ 
as the sum of spectral transforms of individual solitons (normal modes):
\begin{align}
\label{dec0}
\mathcal{S}\{\psi_0\}=\sum_{j=-M}^M\{R(\lambda)=0;\lambda_j,c_j\}=\sum_{j=-M}^M\mathcal{S}\{\phi_j\},
\end{align}
with $\lambda_j$ and $c_j$ given by (\ref{stt}). A remarkable property of integrable PDE
 is that each conserved quantity of the system, such as the energy or the momentum, follows the decomposition (\ref{dec0}) 
 and can be expressed as the sum of conserved quantities of individual solitons. 
 By exploiting this property, we evaluate the Hamiltonian $H$:
\begin{equation}
\label{solh0}
  H=\int_{-\infty}^{\infty}dx\bigg[\bigg\lvert\frac{\partial\psi}{\partial x}\bigg\rvert^2+(\lvert\psi\rvert^2-\rho^2)^2\bigg]=\sum_{j=-M}^M\frac{\nu_j^3}{3},
\end{equation}
whose explicit summation yields $H=\frac{4}{3}\rho^2(\rho^2+1)$, i.e.
the   Hamiltonian of $\psi=\psi_0=-\rho\tanh  x$.  Equation
(\ref{solh0}) states that  the system can be regarded as  a gas of non-interacting particles 
possessing a purely kinetic Hamiltonian of the form $H=\sum_j  m  v_j^2/2$,  
with mass $m=2/3$ and equivalent velocities $v_j=\nu^{3/2}_j$.
The latter governs the observable dynamics  in the ``dual space'' of the wave-function $\psi$,
through Eqs.  (\ref{solnd}-\ref{solndA}).
\newline \indent
The thermodynamics of such soliton gas can be studied by means of \emph{escort distributions} \cite{TOCS}.
Given  an arbitrary measure, whose representative microstates have probability $p_i$, escort
distributions are  constructed  from  the  Lie  transformation  group
$P_k=f(p_k,\beta)=p_k^\beta  /\sum_j  p_j^\beta$.  
The  connection  with statistical  mechanics  is  derived  by  defining the partition function
$Z(\beta)=\sum_j p_j^\beta= \exp(-\Psi)$, correspondingly
$P_i=\exp(\Psi-\beta \mathcal{E}_i)$,  $\mathcal{E}_i=-\ln p_i$ being the energy of the $i-$th microstate ,
and $\beta=1/T>0$ the inverse ``temperature''. 
The Helmholtz free energy  is given by $F(\beta)=\Psi/\beta$. We define the phase-space as the range $[-1,1]$  of  $\lambda/2\rho$.
As a principal  measure  of  the  soliton  gas,  we  consider  its
eigenvalue distribution  $\lambda_i/2\rho$, which is at  the basis of
the  evolution in the  dual  wave-function space through Eq. (\ref{solnd}).  
We then partition the phase-space in $N_\epsilon=2/\epsilon$ boxes of size $\epsilon$, and
define $p_i$ as the normalized ($\sum_j  p_j=1$) probability of finding an eigenvalue 
in the interval $[-1+(i-1)\epsilon, -1+i\epsilon]$ with $i=1,2,...,N_\epsilon$.
Following    the thermostatistics   of   multifractals   \cite{TOCS},  we   study   the
thermodynamic limit of  $F$, defined by 
\begin{equation}
\beta \mathcal{F}=\mathcal{X}=\lim_{V\rightarrow\infty}\frac{\Psi}{V}
\end{equation}
being $V=-\ln\epsilon$ the ``volume''. Correspondingly,
when $\epsilon\rightarrow 0$, the partition function scales as $Z\sim\epsilon^\chi$.
\paragraph{Thermodynamic limit for small particle densities. ---}
We begin by calculating  $\mathcal{X}$ when  the amplitude  $\rho$ is small enough so that the $N_\rho$ eigenvalues $\lambda_j$ are well
separated  in the  spectrum. This situation corresponds  to a  low  density soliton
gas. In this  case, the probability $p_j$ of  finding an eigenvalue in
the   $j-$th   box  of   width   $\epsilon$ is either $0$ or $1/N_\rho$.
Therefore,   in  the   limit $\epsilon\rightarrow   0$, the non-vanishing $p_j$ scales as $\epsilon^0$, 
since  each eigenvalue occupies a single point in the gap $\lvert\lambda/2\rho\rvert\leq 1$.  
The partition function  then reads
\begin{equation}
Z=\sum_j^{N_\epsilon} p_j^\beta=N_\rho \left(\frac{1}{N_\rho}\right)^\beta=N_\rho^{1-\beta} \sim\sum_j^{N_\rho}\epsilon^0\sim
\epsilon^{\mathcal{X}(\beta)}.   
\end{equation}
This   yields    a    very   simple thermodynamics,  where   both  $\mathcal{X}$  and   $\mathcal{F}$  are
zero, with the distribution function $P$ of the eigenvalues, determined
as the limit of $p_j/\epsilon$ for $\epsilon\rightarrow 0$, being a set
of Dirac delta (Fig. \ref{twd}a,b).
The resulting wave evolution is  smooth, without  the  formation  of any  singular
behavior.  We specifically address the  case of
$\rho=2$, where the results of the IST analysis are manageable to
be reported in simple closed form. Equation (\ref{solnd}) yields the particular 3-dark
soliton solution relative to eigenvalues $0,\pm 2\sqrt{3}$ as a function
of $X \equiv 2x, \zeta \equiv 4\sqrt{3} z \in [0,\infty)$:
\begin{equation}
  \psi=\frac{4(\cosh{\zeta}+\cosh{X}+i\sqrt{3}\sinh{\zeta})\sinh{X}}{-3-4\cosh{\zeta}\cosh{X}-\cosh{2X}}.
\end{equation}
Figures \ref{twd}c-e display the level plot evolution of the intensity or density (in BEC) $\lvert\psi\rvert^2$,
and snapshots of $\lvert \psi\rvert^2$ and $S_x=\partial S/\partial x=\mathrm{Im}\{\psi_x/\psi\}$. 
The generation of three dark solitons by $\psi_0$ does not lead to the formation of steep fronts 
during propagation: each soliton slowly splits up from the others throughout the process, 
from the initial overlap state (i.e., at $z=0$ the particles occupy the same position $x=0$) 
to the asymptotic stage where they are well separated. 
As stated above, this relies on the absence of phase transitions in soliton gas (Fig. \ref{twd}a-b), 
owing to a constant free energy landscape $\mathcal{F}=0$. 
In summary, cooperative phenomena and shocks are prohibited at small input amplitudes $\rho$, 
or equivalently, for low density soliton gases where $\mathcal{F}=0$.
\begin{figure}
\includegraphics[width=8.1 cm]{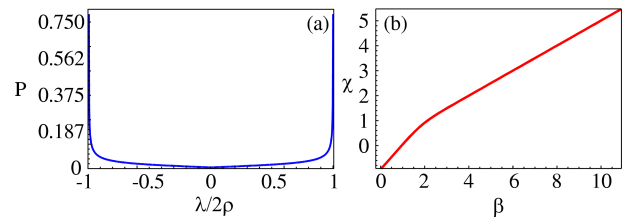}
\caption{\label{pv0}  (Color  online).(a) Eigenvalue measure $P$ (for $\epsilon\rightarrow 0$) versus $\lambda/2\rho$ for a dense gas; 
(b) Free energy $\mathcal{X}$ versus inverse temperature $\beta$.}
\end{figure}
\paragraph{Thermodynamics for dense soliton gases. ---}
When the number of eigenvalues $\lambda$ grows large for $\rho\gg 1$, 
the soliton ensemble turns into a dense gas and the eigenvalue distribution $\lambda/2\rho=\pm\sqrt{1-(1-n/\rho)^2}$ 
becomes a continuous function of the variable $n/\rho$. 
The density of the positive eigenvalues $\mathcal{D}(\lambda/2\rho)$ is readily seen to be $\mathcal{D}(y)=y/2\sqrt{1-y^2}$ and, correspondingly,
it is found that
\begin{equation}
\label{pj}
p_j=\sqrt{1-(j-1)^2\epsilon^2}-\sqrt{1-j^2 \epsilon^2}\text{.}
\end{equation}
For $\epsilon\rightarrow 0$, we identify three different scaling regimes in Eq. (\ref{pj}): (i) $p_j\propto\epsilon^2$ for $\lambda\cong0$; 
(ii) $p_j\propto\epsilon$ for $\lambda\cong\rho$; (iii) $p_j\propto\sqrt{\epsilon}$ for $\lambda\cong2\rho$, with
the partition function $Z$ scaling as
\begin{align}
Z=\sum_j p_j^\beta\sim (N_\epsilon-3)\epsilon^{\beta}+&2\epsilon^{\beta/2}+\epsilon^{2\beta}
\end{align}
with two boxes near $\lambda=\pm 2\rho$ scaling like $\epsilon^{\beta/2}$, one near $\lambda\cong 0$ scaling like $\epsilon^{2\beta}$ and the remaining $N_\epsilon -3$ scaling like $\epsilon^{\beta}$. In the limit $\epsilon\rightarrow 0$, we can actually neglect the contribution of $\epsilon^{2\beta}$ (as the probability $p_j\cong0$ for $\lambda\cong0$). The partition function then reads $Z\sim\epsilon^{\beta-1}+\epsilon^{\beta/2}\sim \epsilon^\mathcal{X}$.
This implies $\mathcal{X}=\mathrm{min}[\beta-1,\beta/2]$ and leads to:
\begin{align}
  \label{free0}
\mathcal{X}=\begin{cases}
\beta-1, &\beta\leq \beta_c\\
\frac{\beta}{2}, &\beta>\beta_c
\end{cases}
\end{align}
being $\beta_c=2$ the critical point such that $\beta_c-1=\beta_c/2$. 
Equation  (\ref{free0}) states that  the system undergoes a  first-order phase  transition as the  inverse temperature
$\beta$ is varied, since  the free energy $\beta \mathcal{F}=\mathcal{X}$  is continuous but  not differentiable
at  $\beta_c$. As  seen  in Fig.  \ref{pv0}b,  $\mathcal{X}$ and $P$ calculated numerically after the expression of the 
eigenvalue $\lambda_n$ support this picture. Such  a phase transition is the main mechanism
leading to shock formation in the dual space of wave dynamics. 
In fact, owing to the different scaling of $p_j$, wave-particles with close
but opposite soliton velocities $\lambda_j$  tend to be accumulated at gap edges;
starting from the initial overlap state, such process generates a
nearly discontinuous variation in the overall frequency $S_x$.  
To  illustrate this dynamics,  we calculate  the  evolution of a dense soliton gas corresponding  to
$\rho=30$  (Fig. \ref{ss0}). A dramatic variation of $S_x$ is observed for  $z=z^*\approx 0.013$ (Fig. \ref{ss0}b), 
with a shock (steep front) in $x=0$ eventually originating from a compressional-like wave ($S_x>0$ for $x<0$ and viceversa),
beyond which fast oscillations in $x$ appear. Each oscillation is indeed associated with a wave particle (dark soliton), 
with solitons splitting up in pairs with opposite velocities around a central black ($\lambda=0$) soliton (Fig. \ref{ss0}a). 
Correspondingly, the intensity $\lvert\psi\rvert^2$ 
 becomes singular in $z^*$ (Fig. \ref{ss0}c). The dynamics of $\psi$ in the dense case ($\rho=30$) is dramatically different if compared to the 3-soliton case (see Figs. \ref{twd}c-d and \ref{ss0}b-c): the absence of cooperative phenomena when the number of soliton is small, in fact, does not lead to any singular behavior (shock wave) in the field propagation along $z$. 
We emphasize that, despite solitons tend to split up from the beginning ($z=0$) due to their different velocities, their collective
behavior results, owing to the compressional wave feature, into a dramatic focusing of the dark notch intensity, 
which turns out to be an easily measurable signature of the critical behavior.
Remarkably this occurs while: (i) no radiation is involved;
(ii) the phase space supporting the gas dynamics is composed by non-interacting particles.
In summary, a dense gas of noninteracting soliton wave-particles turns out to support critical phenomena. 
The latter originates from a singularity of the free-energy, 
which undergoes a metamorphosis as the number of soliton grows and develops a first-order phase transition. 
As shown, the relevant order parameter for the soliton gas transition is the shape of the function $\beta \mathcal{F}$, 
which changes with the input excitation.  

In conclusion, we have illustrated a scenario where the formation of a dispersive shock wave
involves only solitons. In spite of the fact that solitons behave as non-interacting particles with 
well-defined parameters, they do not follow the canonical behavior of systems with non-interacting degrees of freedom. Rather their
cooperation leads to measurable critical phenomena characterized by a free energy that develops a singularity
when the number of solitons grows sufficiently large. This finding can stimulate both novel experiments
and new ideas in the field of statistical mechanics of nonlinear waves.


\end{document}